\documentclass[12pt]{article}
\usepackage{graphics}
\usepackage{amssymb}

\usepackage{bm}% bold math

\newcommand{\rf}[1]{(\ref{#1})}
\newcommand{\beq}{\begin{equation}}

\newcommand{\eeq}{\end{equation}}
\newcommand{\bea}{\begin{eqnarray}}
\newcommand{\eea}{\end{eqnarray}}

\renewcommand{\d}{\mbox{d}}

\renewcommand{\l}{\lambda}

\newcommand{\del}{\delta}

\newcommand{\ra}{\rangle}
\newcommand{\la}{\langle}

\newcommand{\cD}{{\cal D}}

\newcommand{\hH}{{\hat{H}}}

\begin{document}

\begin{center}
\vspace{24pt}
{ \large \bf Planckian Birth of the Quantum de Sitter Universe}

\vspace{30pt}

{\sl J. Ambj\o rn}$\,^{a,c}$
{\sl A. G\"{o}rlich}$\,^{b}$
{\sl J. Jurkiewicz}$\,^{b}$,
and {\sl R. Loll}$\,^{c}$

\vspace{24pt}
{\footnotesize

$^a$~The Niels Bohr Institute, Copenhagen University\\
Blegdamsvej 17, DK-2100 Copenhagen \O , Denmark.\\
{ email: ambjorn@nbi.dk}\\

\vspace{10pt}

$^b$~Institute of Physics, Jagellonian University,\\
Reymonta 4, PL 30-059 Krakow, Poland.\\
{ email: atg@th.if.uj.edu.pl, jurkiewicz@th.if.uj.edu.pl}\\

\vspace{10pt}

$^c$~Institute for Theoretical Physics, Utrecht University, \\
Leuvenlaan 4, NL-3584 CE Utrecht, The Netherlands.\\
{ email: r.loll@phys.uu.nl}\\

\vspace{10pt}
}
\vspace{48pt}

\end{center}

%\addtolength{\baselineskip}{0.20\baselineskip}
%\vspace{2cm}

\begin{center}
{\bf Abstract}
\end{center}
We show that the quantum universe emerging from a nonperturbative, 
Lorentzian sum-over-geometries can be described with high accuracy by a 
four-dimensional de Sitter spacetime.
By a scaling analysis involving Newton's constant, we establish that 
the linear size of the quantum universes under study is in between 
17 and 28 Planck lengths. 
Somewhat surprisingly, the measured quantum fluctuations around the 
de Sitter universe in this regime are to good approximation still 
describable semiclassically. The numerical evidence presented comes 
from a regularization of quantum gravity in terms of causal dynamical 
triangulations.

%\vspace{12pt}
%\noindent
\newpage

\section{\label{intro} Introduction}

To show that the physical spacetime surrounding us can be {\it derived} from 
some fundamental, quantum-dynamical principle is one of the holy grails of 
theoretical physics. The fact that this goal has been eluding us for 
the better 
part of the last half century could be taken as an indication that we have not 
as yet gone far enough in postulating new, exotic ingredients and inventing 
radically new construction principles 
governing physics at the relevant, ultra-high Planckian energy scale. -- 
In this 
letter, we add to previous evidence that such a conclusion may be premature.

Our results are obtained in the framework of 
Lorentzian simplicial quantum gravity, based 
on the concept of causal dynamical triangulations (CDT).
While referring to \cite{blp,4d,al} for details, briefly, 
it defines a nonperturbative way of doing the sum over four-geometries,
assembled from triangular building blocks
such that only causal spacetime histories are 
included. To perform the actual summation, 
one rotates them to spacetimes of Euclidean signature.
The building blocks are four-simplices characterized by 
a cut-off $a$, the side length of the simplices. The continuum limit
of the regularized path integral corresponds to the limit 
$a \to 0$, possibly accompanied by a readjustment of bare 
coupling constants, and such that the physics stays invariant. 
The challenge of a quantum field theory of gravity is to find a theory
which behaves in this way, and suitable observables
to test it. 

How can we judge whether CDT can be taken seriously
as a regularized quantum field theory of gravity?
Our knowledge of the physical world suggests that a viable theory
should {\it generate} a `background geometry' with 
positive cosmological constant, 
superposed with small quantum fluctuations.
The challenge is to obtain this
from a background-independent formulation where no background spacetime
is put in by hand. We have earlier provided indirect
evidence for such a scenario \cite{emerge,semi}. 
Here, we present new computer simulations which confirm this
picture much more directly,  
by establishing the de Sitter nature of the background spacetime, 
quantifying the
fluctuations around it,
and setting a scale for the universes we are dealing with.

\section{Macroscopic de Sitter universe\label{S4}}  

By construction, each path integral geometry (Euclideanized by the
analytic continuation mentioned above) is obtained by gluing 
together four-simplices in a way that respects a global foliation in
discrete proper time $t$. Accordingly, each four-simplex will 
have ``time-like'' and space-like links of length $a$, our short-distance
cutoff.
Spatial slices are compact, three-dimensional 
manifolds of topology $S^3$, glued from space-like
tetrahedra. 
During simulations we fix the time direction 
to a given discrete length $T_{tot}$,
and for reasons of convenience take it to be periodic (with sufficiently
long period so as to not affect results). The bare action used is the 
so-called Regge action, a geometric
realization of the Einstein-Hilbert action on piecewise linear geometries.

A Monte Carlo (MC) simulation of the path integral -- 
for computer-technical reasons 
performed at (almost) fixed four-volume -- will generate a sequence of
spacetime histories, represented by triangulations 
obeying causal gluing rules. 
The key result of CDT consisted in demonstrating that
in a certain range 
of bare coupling constants \footnote{Other choices of bare couplings 
can lead to different and degenerate geometries \cite{blp}.} 
this leads to the {\it emergence of a four-dimensional
universe of well-defined temporal and spatial extension} \cite{emerge}.
Here, we will analyze the geometric nature of this universe in more detail.

Plotting the `shape' of MC-generated path integral configurations in
the form of a time-dependent spatial three-volume $V_3(t)$, one typically 
obtains `blob'-like extended geometries of time extension $T \sim\sqrt[4]{V_4}$.
Individual such configurations are of course not observable, in the same way 
that individual particle paths in a standard path integral are
unphysical. There, one obtains the 
average particle trajectory by taking the average over paths
in the path integral (with a weight dictated by the classical action),
which agrees with the classical particle trajectory
up to corrections of order $\hbar$. Doing this, for instance by MC simulations, 
yields information about both the average trajectory and
the size of the quantum fluctuations, which 
in general will also be of order $\hbar$.

\begin{figure}
\centerline{\scalebox{1.0}{\rotatebox{0}{\includegraphics{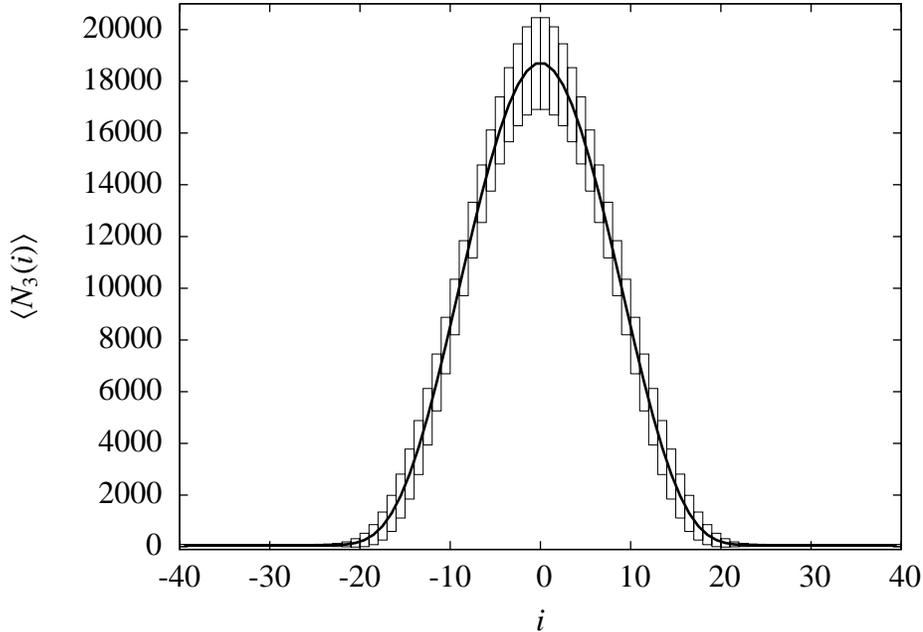}}}}
\caption{\label{fig1} Background geometry $\langle N_3(i)\rangle$: 
MC measurements (for fixed $N_4=362.000$)
and best fit \rf{n1} yield indistinguishable curves at given plot resolution. 
The bars indicate the average size of quantum fluctuations.}
\end{figure}

We will follow the same procedure for our gravitational histories.
Averaging over many different computer-generated configurations will
produce the background geometry around which the quantum universe
fluctuates. Since the ``centre of gravity" of our universes performs a random walk
along the time direction, we take the average $\langle V_3(t)\rangle$ by identifying 
in each history the peak of the volume distribution with $t=0$. ($T_{tot}$ is always
chosen much larger than the time extension $T$ of the universes.)
The results of measuring the average discrete spatial size of the universe at various 
discrete times $i$ are illustrated 
in Fig.\ \ref{fig1} and can be neatly summarized by the formula
\beq\label{n1}
N_3^{cl}(i) = \frac{3}{4} \frac{N_4}{s_0 N_4^{1/4}}  
\cos^3 \left(\frac{i}{s_0 N_4^{1/4}}\right),~~~s_0\approx 0.59,
\eeq
where $N_3(i)$ denotes the number of three-simplices in the spatial slice 
at discretized time $i$ and $N_4$ the 
total number of four-simplices in the entire universe.

We have verified relation \rf{n1} for $N_4$ ranging from 45.500 to 362.000.
The overlap of curves for different $N_4$ according to \rf{n1} constitutes 
a beautiful example of finite-size scaling \cite{next}. Eq.\ \rf{n1} shows that
spatial volumes scale according to $N_4^{3/4}$ and time intervals
according to $N_4^{1/4}$, as one would expect for
a genuinely {\it four}-dimensional spacetime. Translating \rf{n1} to  
a continuum notation leads to 
\beq\label{n2}
\sqrt{g_{tt}}\; V_3^{cl}(t) = V_4 \;\frac{3}{4 B} \cos^3 \left(\frac{t}{B} \right),
\eeq
where we have made the identifications
\beq\label{n3}
V_4 = a^4 C_4 N_4, ~~~~\sqrt{g_{tt}}\;V_3= a^3 C_4 N_3,~~~~t_i = a\; i.
\eeq
In \rf{n3}, $C_4=\sqrt{5}/96$ from the discrete four-simplex volume \cite{4d},
and $\sqrt{g_{tt}}$ is the constant proportionality factor between the time
$t$ and genuine continuum proper time $\tau$, $\tau=\sqrt{g_{tt}}\; t$.
Writing $V_4 =8\pi^2 R^4/{3}$, and $\sqrt{g_{tt}}=R/B$,
eq.\ \rf{n2} is seen to describe a Euclidean {\it de Sitter universe} (a four-sphere, the maximally
symmetric space for positive cosmological constant)
as our searched-for, dynamically generated background geometry!
In the parametrization of \rf{n2} 
this is the classical solution to the action
\beq\label{n5}
S= \frac{1}{24\pi G} \int d t \sqrt{g_{tt}}
\left( \frac{ g^{tt}\dot{V_3}^2(t)}{V_3(t)}+k_2 V_3^{1/3}
-\l V_3(t)\right),
\eeq
where $k_2= 9(2\pi^2)^{2/3}$ and  $\l$ is a Lagrange multiplier,
fixed by requiring that the total four-volume 
be $V_4$, $\int d t \sqrt{g_{tt}} V_3(t) = V_4$. Up to an overall sign, this is precisely 
the Einstein-Hilbert action for the scale factor $a(t)$ of a homogeneous, isotropic universe
(rewritten in terms of $V_3(t) =2\pi^2 a(t)^3$), although we of course never put any such
simplifying symmetry assumptions into the CDT model.
The discretized, dimensionless version of \rf{n5} is
\beq\label{n7b}
S_{discr} =
k_1 \sum_i \left(\frac{(N_3(i+1)-N_3(i))^2}{N_3(i)}+
\tilde{k}_2 N_3^{1/3}\right),
\eeq
where $\tilde{k}_2\propto k_2$. 
The identifications \rf{n3} lead to a
na\"ive continuum limit of the discretized action $S_{discr}$ with
\beq\label{n7c}
G = \frac{a^2}{k_1} \frac{C_4}{24\pi g_{tt} }.
\eeq
Our next aim will be to determine the coefficient $k_1$ in front of the effective 
action \rf{n7b} from the computer simulations. Because of relation \rf{n7c}, this
will give us an estimate of the gravitational coupling constant $G$ in
terms of the lattice spacing $a$.
Since in our units the Planck length is $\ell_{Pl}=\sqrt{G}$, {\it this will set a 
physical length scale for lattice quantities}.
While the classical solution \rf{n2} does not provide information
about the numerical value of $k_1$ in front of \rf{n5}, 
a saddle point calculation shows that the fluctuations
$\delta V_3(t):=V_3(t)-V_3^{cl}(t)$
around $V^{cl}_3(t)= \la V_3(t)\ra $ will be of the order
\beq\label{n7d}
\la (\delta V_3)^2 \ra \sim G \,V_4. 
\eeq
Therefore, {\it if}\hspace{.05cm} it is true that also the fluctuations $\delta N_3(i):=N_3(i) -N_3^{cl}(i)$
of our model are well described by the
mini-superspace action \rf{n5}, we can simply determine $k_1$ from
measuring their correlator.

\section{Fluctuations around de Sitter space\label{fluctuations}}

Having demonstrated that the action \rf{n5} 
gives a perfect description of the measured
$V_3^{cl}(t)$, we will now show that {\it it also describes the observed 
quantum fluctuations around
de Sitter space}, defined by the correlator 
\beq\label{n7} 
C(t,t')= \la \delta V_3(t) \delta V_3 (t') \ra.
\eeq
\begin{figure}
\centerline{{\scalebox{1.0}{\rotatebox{0}{\includegraphics{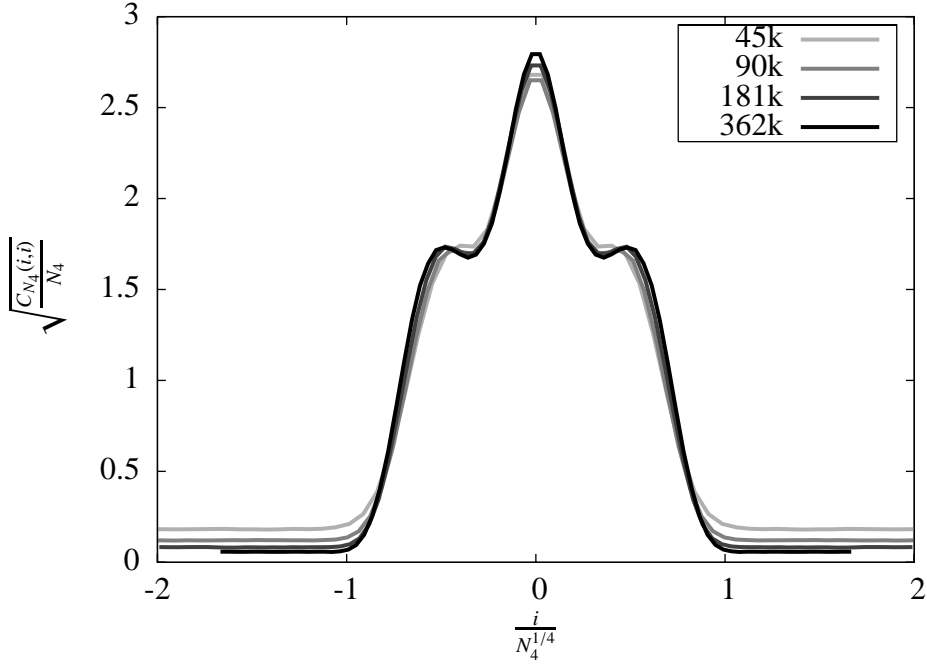}}}}}
\caption{\label{fig0} Analyzing the quantum fluctuations of Fig.\ \ref{fig1}: 
diagonal entries $F(t,t)^{1/2}$ of the scaling function 
$F$ from \rf{n7f}, for $N_4=$ 45.500,
91.000, 181.000 and 362.000.}
\end{figure}
Our first observation from the data is that the discretized version of \rf{n7}
scales with the four-volume according to
\beq
C_{N_4}(i,i') = \la \delta N_3(i) \delta N_3 (i')\ra =
N_4 \; F\Big({i}/N^{1/4}_4,{i'}/N_4^{1/4}\Big), \label{n7f}
\eeq
where $F$ is a universal scaling function of order $a^0$. 
This is illustrated by Fig.\ \ref{fig0} for $C_{N_4}^{1/2}(i,i)$, 
corresponding precisely to the fluctuations $\la (\del V_3(t))^2\ra^{1/2}$ 
of Fig.\ \ref{fig1}.
This scaling implies that the action \rf{n7b} can only
describe the fluctuations measured for different $N_4$ if
$k_1$ is independent of $N_4$, thus confirming the scaling behaviour 
$G \sim a^2$ anticipated in \rf{n7c}.  

To demonstrate that $F(t,t')$ is indeed described by the
effective actions \rf{n5}, \rf{n7b}, let us for convenience adopt a 
continuum language and compute its expected behaviour.
Expanding \rf{n5} around the classical solution 
as $V_3(t) = V_3^{cl}(t) + x(t)$,
the quadratic fluctuations are given by
\bea
\la x(t) x(t')\ra &\! =\! & 
\int \cD x(s)\; x(t)x(t')\; 
e^{ -\frac{1}{2}\int\!\!\int d s d s' x(s) M(s,s') x(s')} \nonumber\\
&\! =\! &  M^{-1}(t,t'),\label{n7a}
\eea
where $\cD x(s)$ is the normalized measure and 
the quadratic form $M(t,t')$ is determined by expanding the 
effective action $S$ to second order in $x(t)$, 
\beq\label{n8}
S(V_3) = S(V_3^{cl}) + \frac{1}{18\pi G}\frac{B}{V_4}  \int \d t \; x(t) \hH 
 x(t).
\eeq
In \rf{n8}, $\hH$ denotes the Hermitian operator
\beq\label{n9}
\hH= -\frac{\d }{\d t} \frac{1}{\cos^3 (t/B)}\frac{\d }{\d t} -
\frac{4}{B^2\cos^5(t/B)},
\eeq
which must be diagonalized 
under the constraint that $\int dt \sqrt{g_{tt}}\; x(t) =0$, since
$V_4$ is kept constant.

\begin{figure}
\centerline{{\scalebox{1.0}{\rotatebox{0}{\includegraphics{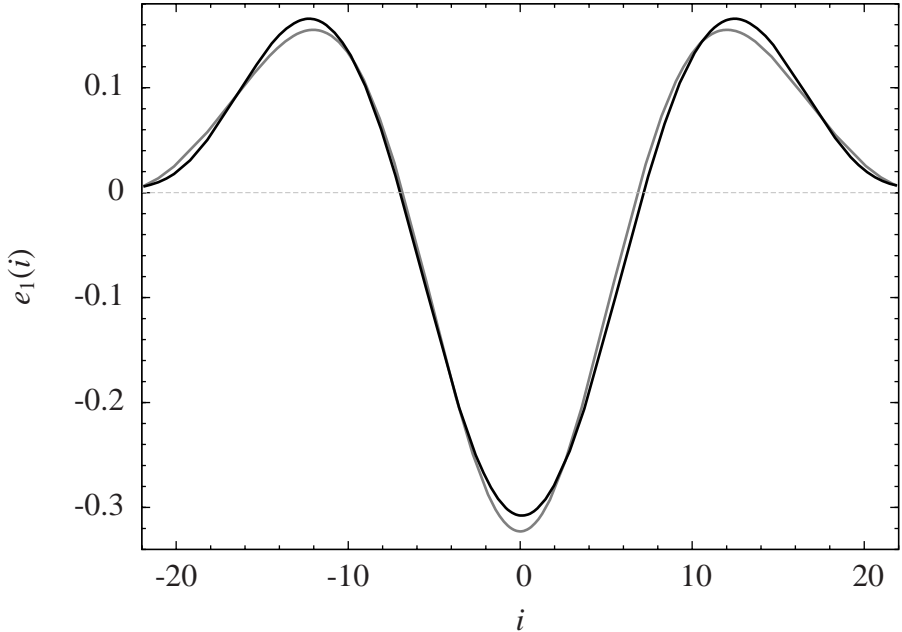}}}}}
\caption{\label{fig2}Comparing the highest eigenvector of $C(t,t')$
and the lowest eigenvector of $M^{-1}(t,t')$. }
\end{figure}

Let $e_n(t)$ be the eigenfunctions of the quadratic form 
given by \rf{n8} with the volume constraint enforced, 
ordered according to increasing eigenvalues $\l_n$.
If this cosmological continuum model were to give the correct 
description of the computer-generated universe, the matrix 
\beq\label{n12}
M^{-1}(t,t') = \sum_{n=1}^\infty \frac{e_n(t)e_n(t')}{\l_n}.
\eeq 
should be proportional to the measured correlator $C(t,t')$.
Fig.\ \ref{fig2} shows the highest eigenfunction calculated
from the data, the matrix $C(t,t')$, 
and the corresponding lowest eigenfunction 
calculated from the effective action, the matrix $M(t,t')$. 
The agreement is very good, in particular taking into account that
no parameter is adjusted in the action (we simply take 
$B=s_0N_4^{1/4}$ in \rf{n2} and \rf{n8}, i.e. 14.47$a$
for $N_4 = 362.000$). One can also compare the data and 
the matrix $M^{-1}(t,t')$ calculated from \rf{n12} directly. The agreement
is again good, although less spectacular than
in Fig.\ \ref{fig2}. 
Awaiting publication of a full analysis of the data \cite{next}, suffice it to
say that within measuring accuracy the fluctuations of the spatial volume 
are described by the mini-superspace action \rf{n5}.  
This enables us to numerically estimate $k_1\approx 0.016$, which according
to \rf{n7c} leads to $G\approx 0.22 a^2$, or $\ell_{Pl}\approx 0.47 a$. 
In other words, {\it the linear size of the quantum de Sitter universes 
studied here lies in the range of 17-28 Planck lengths}.

\section{Discussion\label{discuss}}

The CDT model of quantum gravity is extremely simple, namely, the path integral 
over the class of causal geometries with a global time foliation. In order
to perform this summation explicitly, we introduce a grid of piecewise linear 
geometries, much in the same way as when defining the path integral in 
quantum mechanics. Next, we rotate each of these geometries to Euclidean 
signature and use as bare action the Einstein-Hilbert action \footnote{Of course, the
full, effective action, including measure contributions, will contain all higher-derivative 
terms.} in Regge form. Nothing else is put in.

The resulting superposition exhibits scaling behaviour as function of the 
four-volume, and we observe the appearance of a well-defined average 
geometry, that of de Sitter space.  
We are definitely in a quantum regime,
since the fluctuations around de Sitter space are sizeable, as can be seen in 
Fig.\ \ref{fig1}. Both the average geometry and the quantum fluctuations are 
well described by the mini-superspace action \rf{n5}. Unlike in standard 
cosmological treatments, this description is the {\it outcome} of a 
nonperturbative evaluation
of the {\it full} path integral, with everything but the scale factor (equivalently, $V_3(t)$)
summed over. Measuring the correlations of the fluctuations in the computer
simulation enabled us to determine the continuum gravitational coupling
constant to $G\approx 0.22 a^2$, thereby introducing an absolute physical 
length scale. Within measuring accuracy, our de Sitter universes (with volumes
in the range 22.000-173.000 $\ell_{Pl}^4$) are seen to behave perfectly 
semiclassically. 

Can we study smaller universes, which are themselves of Planck size? 
Taking the coupling $G$ as a true measure of the gravitational
coupling constant, the simplest way is as follows.
We are free to vary $N_4$ and the bare gravitational coupling 
constant $g_0$ of the Regge action (see \cite{blp}
for further details on the bare coupling constants), with the effective constant $k_1$
a function of $g_0$. If we adjusted $g_0$ such that in the limit $N_4 \to \infty$ both
\beq\label{n100}
V_4 \sim N_4 a^4~~~{\rm and}~~~G\sim a^2/k_1(g_0)
\eeq
remained constant (i.e.\ $k_1(g_0) \sim 1/\sqrt{N_4}$), we would eventually 
penetrate into the (sub-)Planckian regime. Since we have already seen deviations 
from classicality at short scales \cite{blp}, 
we would expect the canonical scaling
of $G$ to change there, or, stated differently, the simple 
effective action \rf{n5} to be no longer valid. 
Renormalization group methods have produced 
predictions for the scaling violations of $G$ in the 
context of asymptotic safety \cite{reuteretc}, which
in principle we should be able to test. 
In this context it would be ideal to have an
observable with an associated correlation length
that could be kept constant when expressed in terms of $V_4^{1/4}$.
A further step will be to include matter in the model and verify directly that 
$G$ can indeed be interpreted as Newton's constant, perhaps
along the lines pursued earlier in Euclidean 
quantum gravity \cite{js}. All of these issues are currently under investigation.

\subsection*{Acknowledgments}
All authors acknowledge support by
ENRAGE (European Network on
Random Geometry), a Marie Curie Research Training Network, 
contract MRTN-CT-2004-005616, and 
A.G. and J.J. by COCOS (Correlations in Complex Systems), a Marie Curie Transfer
of Knowledge Project, contract MTKD-CT-2004-517186, both in the 
European Community's Sixth Framework Programme.
JJ acknowledges a partial support by the Polish Ministry of Science and
Information Technologies grant 1P03B04029 (2005-2008).


\begin{thebibliography}{99}

%\cite{Ambjorn:2005qt}
\bibitem{blp}
  J.~Ambj\o rn, J.~Jurkiewicz and R.~Loll,
  %{\it Reconstructing the universe,}
  Phys.\ Rev.\  D {\bf 72} (2005) 064014
  [arXiv:hep-th/0505154].
  %%CITATION = PHRVA,D72,064014;%%

%\cite{Ambjorn:2001cv}
\bibitem{4d}
  J.~Ambj\o rn, J.~Jurkiewicz and R.~Loll,
%{\it Dynamically triangulating Lorentzian quantum gravity,}
  Nucl.\ Phys.\  B {\bf 610} (2001) 347
  [arXiv:hep-th/0105267].
  %%CITATION = NUPHA,B610,347;%%


%\cite{Ambjorn:1998xu}
\bibitem{al}
  J.~Ambj\o rn and R.~Loll,
  %{\it Non-perturbative Lorentzian quantum gravity, causality and topology
%  change,}
  Nucl.\ Phys.\  B {\bf 536} (1998) 407
  [arXiv:hep-th/9805108].
  %%CITATION = NUPHA,B536,407;%%

%\cite{Ambjorn:2004qm}
\bibitem{emerge}
  J.~Ambj\o rn, J.~Jurkiewicz and R.~Loll,
  %{\it Emergence of a 4D world from causal quantum gravity,}
  Phys.\ Rev.\ Lett.\  {\bf 93} (2004) 131301
  [arXiv:hep-th/0404156].
  %%CITATION = PRLTA,93,131301;%%

%\cite{Ambjorn:2004pw}
\bibitem{semi}
  J.~Ambj\o rn, J.~Jurkiewicz and R.~Loll,
  %{\it Semiclassical universe from first principles,}
  Phys.\ Lett.\  B {\bf 607} (2005) 205
  [arXiv:hep-th/0411152].
  %%CITATION = PHLTA,B607,205;%%

\bibitem{next}
J.~Ambj\o rn, A.\ G\"orlich, J.~Jurkiewicz and R.~Loll, to appear.

\bibitem{reuteretc}
%\bibitem{Reuter:2007rv}
  M.~Reuter and F.~Saueressig,
  %``Functional Renormalization Group Equations, Asymptotic Safety, and Quantum
  %Einstein Gravity,''
  arXiv:0708.1317 [hep-th].\\
  %%CITATION = ARXIV:0708.1317;%%
%\cite{Litim:2003vp}
%\bibitem{Litim:2003vp}
  D.~F.~Litim,
  %``Fixed points of quantum gravity,''
  Phys.\ Rev.\ Lett.\  {\bf 92} (2004) 201301
  [arXiv:hep-th/0312114].\\
  %%CITATION = PRLTA,92,201301;%%
%\cite{Reuter:2007rv}
%\cite{Codello:2006in}
%\bibitem{Codello:2006in}
  A.~Codello and R.~Percacci,
  %``Fixed Points of Higher Derivative Gravity,''
  Phys.\ Rev.\ Lett.\  {\bf 97} (2006) 221301
  [arXiv:hep-th/0607128].\\
  %%CITATION = PRLTA,97,221301;%%
%\bibitem{kawai}
H.~Kawai, Y.~Kitazawa and M.~Ninomiya,
%{\it Renormalizability of quantum gravity near two dimensions,}
  Nucl.\ Phys.\ B\ 467 (1996) 313-331 [arXiv:hep-th/9511217].\\
%%CITATION = HEP-TH 9511217;%%
%\cite{Niedermaier:2006ns}
%\bibitem{Niedermaier:2006ns}
  M.~Niedermaier,
  %``The asymptotic safety scenario in quantum gravity: An introduction,''
  Class.\ Quant.\ Grav.\  {\bf 24} (2007) R171
  [arXiv:gr-qc/0610018].\\
  %%CITATION = CQGRD,24,R171;%%
%\cite{Hamber:2005dw}
%\bibitem{Hamber:2005dw}
  H.~W.~Hamber and R.~M.~Williams,
  %``Nonlocal effective gravitational field equations and the running of
  %Newton's G,''
  Phys.\ Rev.\  D {\bf 72} (2005) 044026
  [arXiv:hep-th/0507017].
  %%CITATION = PHRVA,D72,044026;%%

%\cite{deBakker:1996qf}
\bibitem{js}
  B.~V.~de Bakker and J.~Smit,
  %``Gravitational binding in 4D dynamical triangulation,''
  Nucl.\ Phys.\  B {\bf 484} (1997) 476
  [arXiv:hep-lat/9604023].
  %%CITATION = NUPHA,B484,476;%%

\end{thebibliography}
\end{document}